\documentclass[10pt,a4paper]{article}
\usepackage{jheppub_kim}
\usepackage{pdflscape}
\usepackage{amsmath}
\usepackage{amssymb}
\usepackage{dcolumn}
\usepackage{bm}
\usepackage{color}
\usepackage{epsfig}
\usepackage{amsfonts}
\usepackage{graphicx}

\usepackage{graphicx}
\usepackage{dcolumn}
\usepackage{bm}

\usepackage{hyperref} 
\begin{document}
 \title{Static Magnetic Brane Solutions in Quartic Quasi-topological Gravity with Power-Law Maxwell Nonlinear Electrodynamics}

\author[a]{Alireza Alizadeh,}
 \author[a]{Saheb Soroushfar,\footnote{Corresponding Author}}
  \author[b]{Mohammad Ghanaatian,}
 \author[c]{Sayyed Mehrab Ramezani,}
\author[d,e]{Sudhaker Upadhyay,\footnote{Visiting Associate, Inter-University Centre for Astronomy and Astrophysics (IUCAA) Pune-411007, Maharashtra, India}}
\author[f]{Kairat Myrzakulov}
\affiliation[a]{Department of Physics, College of Sciences, Yasouj University, 7591874934, Yasouj, Iran}
\affiliation[b]{Department of Physics, Jahrom University, 7413766171 Jahrom, Iran}
\affiliation[c]{Department of Mathematics, College of Sciences, Yasouj University, 7591874934, Yasouj, Iran}
\affiliation[d] {Department of Physics, K.L.S. College, Magadh University, Nawada-805110, Bihar,  India}
\affiliation[e] {School of Physics, Damghan University, Damghan, 3671641167, Iran.}
\affiliation[f]{Department of General \& Theoretical Physics, L. N. 
Gumilyov Eurasian National University, Astana, 010008, Kazakhstan}

\emailAdd{Alirezaalizadeh02@gmail.com}
\emailAdd{soroush@yu.ac.ir}
\emailAdd{dr.ghanaatian@yahoo.com}
\emailAdd{m.ramezani@yu.ac.ir}
\emailAdd{sudhakerupadhyay@gmail.com}
\emailAdd{krmyrzakulov@gmail.com}
  
\abstract{ 
In this paper, we investigate static magnetic brane solutions in
$(n+1)$-dimensional quartic quasi-topological gravity coupled to power-law
Maxwell nonlinear electrodynamics. We derive the corresponding solutions
and analyze their geometric and physical properties. We observe that the
solutions are horizonless and free of curvature singularities in the
physically accessible region, while the spacetime generally exhibits a
conical singularity with a nonzero angular deficit at the lower radial
endpoint. We demonstrate that the endpoint is independent of the
higher-curvature couplings $\lambda$, $\mu$, and $c$, whereas the
near-endpoint geometry is mainly controlled by the magnetic parameter $q$
and the nonlinear parameter $s$. At large radial distances, we find that
the higher-curvature couplings govern the asymptotic behavior of the
metric. We also establish that finite asymptotic solutions require
$1/2<s\leq2$ and analyze the dependence of the angular deficit on $q$,
$s$, and the endpoint parameter $\rho_{+}$. Using the counterterm method,
we calculate the conserved mass density and show that the static magnetic
brane carries no net electric charge. Finally, we verify that the
power-law Maxwell theory recovers linear Maxwell electrodynamics at
$s=1$ and Einstein gravity when $\lambda=\mu=c=0$.
 
 }

\maketitle
 
\section{Introduction}
Generalized theories of gravity and nonlinear electrodynamics are of significant importance, as they offer potential solutions to numerous physical problems. Generalized gravity theories can provide alternatives to standard gravity, particularly in addressing phenomena such as dark energy, and can describe various cosmological phases \cite{1,2}. Furthermore, some physical systems in nature exhibit field equations that are inherently nonlinear, a characteristic also found in gravitational systems. Modified gravity models show promise in unifying early-time and late-time inflation \cite{3,4,5,45}.

We consider a novel gravitational action that combines some of the simplicity and beneficial properties of Lovelock actions with field equations in lower dimensions (e.g., five dimensions) that are not immediately apparent. Consequently, a new class of generalized gravity, termed quasi-topological gravity, was introduced. It shares similarities with Lovelock gravity but possesses improved characteristics. Quasi-topological gravity incorporates cubic and quartic terms of the Riemann tensor and imposes no dimensional limitations beyond five \cite{6,7,8}. 

Essentially, because Einstein's field equations are not the most comprehensive higher-dimensional equations and may not fully satisfy Einstein's assumptions, quasi-topological gravity emerges as a higher-derivative theory capable of resolving these issues \cite{9}. Nonlinear electrodynamics has also found applications in cosmological models, such as describing the inflationary epoch and the late-time accelerated expansion of the universe. Some applications of nonlinear electrodynamics have been investigated \cite{10}. This theory has also been instrumental in discovering the first exact regular black hole solutions sourced by nonlinear electrodynamics that satisfy the weak energy condition \cite{11}. These compelling reasons motivate our investigation into nonlinear electrodynamics.

On the other hand, there are several motivations for studying nonlinear electrodynamics. This theory can eliminate or mitigate the singularity of the electric field associated with a point charge at the origin, a known issue in linear Maxwell theory (one of the most successful linear formulations of electromagnetism). Another crucial reason for the development of nonlinear electrodynamics is its capacity to describe complex systems, chaotic phenomena, as well as the behavior of compact astrophysical objects such as neutron stars and pulsars. Moreover, it aligns with the AdS/CFT correspondence and string theory frameworks, playing a role in the description of pair creation in Hawking radiation \cite{12,13,14}.

The initial efforts to establish a link between nonlinear electrodynamics and gravity were undertaken by Hofmann \cite{15}. Nonlinear electrodynamic field theory was introduced by Born-Infeld \cite{16,17} through various Lagrangians, including logarithmic and exponential forms \cite{41,18,19,20,21}. Subsequently, a new nonlinear Lagrangian known as the power-law Maxwell Lagrangian was introduced. In addition to the aforementioned advantages, the power-law Maxwell Lagrangian offers an additional benefit over the first three Lagrangians: it can establish conformal invariance in higher dimensions \cite{22,23,42,44}. 

This theory is defined as follows. Four nonlinear Lagrangians have been proposed to address the aforementioned problems: to solve the electron self-energy problem, the nonlinear electrodynamic Lagrangian is given by
\begin{equation}
\mathcal{L}_{\rm NLED}
=
\mathcal{F}_{m}^{\,s}.
\label{2.16}
\end{equation}
Here, $s$ is the nonlinear parameter of the power-law Maxwell theory, and $\mathcal{F}_{m} = F_{\mu\nu}F^{\mu\nu}$, where $F_{\mu\nu}$ is the electromagnetic field tensor, which is defined as $F_{\mu\nu} = \partial_{\mu}A_{\nu} - \partial_{\nu}A_{\mu}$, and $A_{\mu}$ is the vector potential.

 Unlike exponential, logarithmic, or Born–Infeld electrodynamics, which are primarily motivated by string theory or electron self-energy regularization, the power-law Maxwell Lagrangian  offers a unique and physically motivated advantage: it restores conformal invariance in higher dimensions for a specific critical power \(s=(n+1)/4\) \cite{23}. This property makes it particularly suitable for studying scale-invariant regimes in the AdS/CFT correspondence and for constructing exact magnetic brane solutions with tunable asymptotic behavior. Furthermore, while Born–Infeld  theory  and its generalizations reduce to linear Maxwell theory only in the weak-field limit, the power-law model does so exactly as \(s \to 1\), providing a controlled deformation of the standard theory. Its field equations remain algebraic and analytically tractable even when coupled to quartic quasi-topological gravity, allowing for explicit control over the deficit angle and the mass parameter without relying on numerical integration. In this sense, the present model is not a mere mathematical generalization, but a physically motivated laboratory for probing the interplay between conformal symmetry, higher-curvature corrections, and nonlinear electromagnetic sources in horizonless spacetimes.

In recent years, magnetic branes within cubic quasi-topological gravity, in the presence of Maxwell and Born-Infeld electromagnetic fields, have been investigated. Similarly, quasi-topological magnetic branes coupled to nonlinear electrodynamics such as exponential and logarithmic forms have also been studied. The solutions derived for this magnetic brane are horizonless and have no curvature \cite{21,24}. Also, the study of quintic quasi-topological gravity and thermodynamics of quasi-topological magnetic branes coupled to nonlinear electrodynamics has been conducted \cite{25}. 

Building upon this, we aim to advance this research by studying magnetic brane solutions with power-law Maxwell nonlinear electrodynamics in quartic quasi-topological gravity. Extraction of cubic quasi-topological black hole solutions coupled to power-law Maxwell nonlinear electrodynamics has been performed \cite{23}. Investigation of thermodynamics and multi-horizon solutions in quartic quasi-topological gravity in the presence of power-law Maxwell nonlinear electrodynamics has been carried out \cite{26}. Thermodynamics of static solutions in (n+1)-dimensional quintic quasi-topological gravity has been studied \cite{27}. Also, charged dilaton black holes in the presence of nonlinear power-Maxwell electrodynamics have been investigated \cite{28}.

In this paper, we consider static magnetic brane solutions in quartic
quasi-topological gravity coupled to power-law Maxwell nonlinear
electrodynamics. We further derive the corresponding solutions and analyze their
geometric and physical properties, including their horizon structure,
curvature behavior, conical singularity, asymptotic properties, and
conserved quantities.  

The structure of this paper is organized as follows:  
In Sec. \ref{sec2}, we introduce  the metric of the horizonless spacetime and the action incorporating nonlinear electrodynamics and quartic quasi-topological terms, followed by the derivation of the field equations and solutions. In Sec. \ref{sec3}, we  analyze  the physical properties and behavior of the obtained solutions, supported by illustrative figures. We discuss  the counterterm method to compute the conserved quantities of the spacetime in Sec. \ref{sec4}. Finally, in Sec. \ref{sec5}, we summarize  the findings and presents the overall conclusions of the paper. 
\section{Static Magnetic Brane Solutions}\label{sec2}

We consider a static magnetic brane solution in $(n+1)$-dimensional
quartic quasi-topological gravity coupled to power-law Maxwell
nonlinear electrodynamics. The gravitational and electromagnetic
dynamics are governed by the action
\begin{equation}
I_G=
\frac{1}{16\pi}
\int d^{n+1}x\,\sqrt{-g}\,
\left[
-2\Lambda
+\mathcal{L}_1
+\lambda\mathcal{L}_2
+\mu\mathcal{L}_3
+c\mathcal{L}_4
+\mathcal{L}_{\rm NLED}
\right],
\label{eq:action}
\end{equation}
where
\begin{equation}
\mathcal{L}_1=R,
\end{equation}
and
\begin{equation}
\mathcal{L}_2=
R_{\mu\nu\rho\sigma}R^{\mu\nu\rho\sigma}
-4R_{\mu\nu}R^{\mu\nu}
+R^2
\end{equation}
is the Gauss--Bonnet invariant. The quantities
$\mathcal{L}_3$ and $\mathcal{L}_4$ denote the cubic and quartic
quasi-topological curvature invariants, respectively, given by 
\begin{eqnarray}
\mathcal{L}_{3} &=& R_{ab}^{cd}R_{cd}^{ef}R_{ef}^{ab} + \frac{1}{(2n-1)(n-3)} \cdot \frac{3(3n-5)}{8} R_{abcd}R^{abcd}R \nonumber \\
&-& 3(n-1)R_{abcd}R_{e}^{abc}R^{de} + 3(n+1)R_{abcd}R^{ac}R^{bd} + 6(n-1)R_{a}^{b}R_{b}^{c}R_{c}^{a} \nonumber \\
&-& \frac{3(3n-1)}{2}R_{a}^{b}R_{b}^{a}R + \frac{3(n+1)}{8}R^{3},\label{6}\\
\mathcal{L}_{4} &=& c_{1}R_{abcd}R^{cdef}R_{ef}^{hg}R_{hg}^{ab} + c_{2}R_{abcd}R^{abcd}R_{ef}R^{ef} + c_{3}RR_{ab}R^{ac}R_{c}^{b} \nonumber \\
&+ &c_{4}\left(R_{abcd}R^{abcd}\right)^{2} + c_{5}R_{ab}R^{ac}R_{cd}R^{db} + c_{6}RR_{abcd}R^{ac}R^{db} \nonumber \\
&+& c_{7}R_{abcd}R^{ac}R^{be}R_{e}^{d} + c_{8}R_{abcd}R^{acef}R_{e}^{b}R_{f}^{d} + c_{9}R_{abcd}R^{ac}R_{ef}R^{bedf} + c_{10}R^{4} \nonumber \\
&+& c_{11}R_{abcd}R^{abcd}R^{2} + c_{12}R_{ab}R^{ab}R^{2} + c_{13}R_{abcd}R^{cbef}R_{efg}^{c}R^{dg} \nonumber \\
&+ &c_{14}R_{abcd}R^{acef}R_{gehf}R^{gbhd}.  \label{7}
\end{eqnarray}
Here, 
\begin{align}
c_{1} &= -(n-1)\left(n^{7} - 3n^{6} - 29n^{5} + 170n^{4} - 349n^{3} + 348n^{2} - 180n + 36\right), \nonumber \\
c_{2} &= -4(n-3)\left(2n^{6} - 20n^{5} + 65n^{4} - 81n^{3} + 13n^{2} + 45n - 18\right), \nonumber \\
c_{3} &= -64(n-1)(3n^{2} - 8n + 3)(n^{2} - 3n + 3), \nonumber \\
c_{4} &= -\left(n^{8} - 6n^{7} + 12n^{6} - 22n^{5} + 114n^{4} - 345n^{3} + 468n^{2} - 270n + 54\right), \nonumber \\
c_{5} &= 16(n-1)\left(10n^{4} - 51n^{3} + 93n^{2} - 72n + 18\right), \nonumber \\
c_{6} &= 32(n-1)^{2}(n-3)^{2}(3n^{2} - 8n + 3), \nonumber \\
c_{7} &= 64(n-2)(n-1)^{2}(4n^{3} - 18n^{2} + 27n - 9), \nonumber \\
c_{8} &= -96(n-1)(n-2)(2n^{4} - 7n^{3} + 4n^{2} + 6n - 3), \nonumber \\
c_{9} &= 16(n-1)^{3}(2n^{4} - 26n^{3} + 93n^{2} - 117n + 36), \nonumber \\
c_{10} &= n^{5} - 31n^{4} + 168n^{3} - 360n^{2} + 330n - 90, \nonumber \\
c_{11} &= 2\left(6n^{6} - 67n^{5} + 311n^{4} - 742n^{3} + 936n^{2} - 576n + 126\right), \nonumber \\
c_{12} &= 8\left(7n^{5} - 47n^{4} + 121n^{3} - 141n^{2} + 63n - 9\right), \nonumber \\
c_{13} &= 16n(n-1)(n-2)(n-3)(3n^{2} - 8n + 3), \nonumber \\
c_{14} &= 8(n-1)\left(n^{7} - 4n^{6} - 15n^{5} + 122n^{4} - 287n^{3} + 297n^{2} - 126n + 18\right), \label{8}
\end{align}
We consider the static metric ansatz
 \begin{equation}
ds^2=
-\frac{\rho^2}{l^2}dt^2
+\frac{d\rho^2}{f(\rho)}
+l^2g(\rho)d\phi^2
+\frac{\rho^2}{l^2}dX^2 ,
\label{2.11}
\end{equation}
where
\begin{equation}
dX^2=\sum_{i=1}^{n-2}(dx^i)^2 ,
\label{2.12}
\end{equation}
and $\phi$ is a dimensionless angular coordinate with
$0\leq\phi\leq2\pi$.

For the magnetic configuration, we choose the vector potential as
\begin{equation}
A_\mu=h(\rho)\delta^\phi_\mu ,
\label{2.13}
\end{equation}
so that the only nonvanishing components of the field-strength tensor
are
\begin{equation}
F_{\rho\phi}=-F_{\phi\rho}=h'(\rho),
\label{2.14}
\end{equation}
Now, we introduce the auxiliary functions
\begin{equation}
\psi(\rho)
=
-\frac{l^2f(\rho)}{\rho^2},
\label{2.17}
\end{equation} 
\begin{equation}
g(\rho)=N^2(\rho)f(\rho).
\label{2.18}
\end{equation}
After substituting the metric and electromagnetic ansatz into the
action (\ref{eq:action}) and integrating by parts, the reduced action can be written as
\begin{equation}
I_{\rm red}
=
\frac{n-1}{16\pi l^2}
\int d^nx
\int d\rho\,
N(\rho)
\left[
\rho^n\mathcal{P}'(\rho)
+
\eta_s
\rho^{n-1}
N^{-2s}(\rho)
h'^{\,2s}(\rho)
\right],
\label{2.19}
\end{equation}
where
\begin{equation}
\mathcal{P}(\rho)
=
1+\psi+\lambda\psi^2+\mu\psi^3+c\psi^4,
\label{2.20}
\end{equation}
and
\begin{equation}
\eta_s =
\frac{2^s l^{\,2-2s}}{n-1}.
\label{2.21}
\end{equation}
Upon variation of the reduced action with respect to $\psi$, we get
\begin{equation}
\left(
1+2\lambda\psi+3\mu\psi^2+4c\psi^3
\right)N'(\rho)=0.
\label{2.22}
\end{equation}
For the nondegenerate branch  $
N'(\rho)=0$. 
The constant value of $N$ can be absorbed into a rescaling of the
time and angular coordinates. We therefore choose
$
N(\rho)=1,$ 
which implies
\begin{equation}
g(\rho)=f(\rho).
\label{2.25}
\end{equation}
The variation of the reduced action with respect to $h(\rho)$ gives
\begin{equation}
\frac{d}{d\rho}
\left[
\rho^{n-1}h'^{\,2s-1}
\right]
=0.
\label{2.26}
\end{equation}
Here, an irrelevant constant multiplicative factor has been absorbed
into the definition of the integration constant.
Thus,
\begin{equation}
\rho^{n-1}h'^{\,2s-1}
=
\mathcal{Q}_{s,n},
\label{2.27}
\end{equation}
where $\mathcal{Q}_{s,n}$ is a constant related to the magnetic charge
parameter $q$. To recover the Maxwell limit
 at $s=1$, we normalize this constant. Therefore, the resulting magnetic field can be written as
\begin{equation}
F_{\phi\rho}
=
-h'(\rho)
=
\frac{s}{2q}\,
\rho^{n-1}
\left(
\frac{2q^2l^{\,n-2}}
{s^2\rho^{2n-2}}
\right)^{\frac{s}{2s-1}} =\mathcal{A}_{s,n}\,
\rho^{-\frac{n-1}{2s-1}},
\label{2.28}
\end{equation}
where
\begin{equation}
\mathcal{A}_{s,n}
=
\frac{s}{2q}
\left(
\frac{2q^2l^{\,n-2}}{s^2}
\right)^{\frac{s}{2s-1}}.
\label{2.30}
\end{equation}
The radial dependence of the magnetic field is therefore
$
F_{\phi\rho}
\propto
\rho^{-\frac{n-1}{2s-1}}$.
 In the linear Maxwell limit, $s=1$, Eq.~(\ref{2.28}) becomes
\begin{equation}
F_{\phi\rho}
=
\frac{q\,l^{\,n-2}}{\rho^{n-1}},
\qquad s=1,
\label{2.32}
\end{equation}
which is the standard magnetic field profile in $(n+1)$ dimensions.
Since $
F_{\phi\rho}
=
-\partial_\rho A_\phi$,
 the vector potential follows by integration. For
$2s\neq n$, we obtain
\begin{equation}
A_\phi(\rho)
=
A_{s,n}\,
\rho^{\frac{2s-n}{2s-1}},
\qquad
2s\neq n,
\label{2.34}
\end{equation}
where
\begin{equation}
A_{s,n}
=
\frac{2s-1}{n-2s}
\frac{s}{2q}
\left(
\frac{2q^2l^{\,n-2}}{s^2}
\right)^{\frac{s}{2s-1}} .
\label{2.35}
\end{equation}
An arbitrary additive constant has been omitted because it can be
removed by a gauge transformation.
When $
2s=n,$ 
the power-law expression becomes logarithmic and in this case,
 \begin{equation}
A_\phi(\rho)
=
-Q_n\ln\rho ,
\qquad
s=\frac{n}{2},
\label{2.38}
\end{equation}
where
\begin{equation}
Q_n
=
\frac{n}{4q}
\left(
\frac{8q^2l^{\,n-2}}{n^2}
\right)^{\frac{n}{2(n-1)}} .
\label{2.39}
\end{equation}
Thus, the commonly used logarithmic case $s=2$ corresponds specifically
to $n=4$. 

Now, variation of the reduced
action with respect to $N(\rho)$, followed by setting $N=1$, gives
\begin{equation}
\rho^n\mathcal{P}'(\rho)
-
(2s-1)\eta_s
\rho^{n-1}h'^{\,2s}
=0.
\label{2.41}
\end{equation}
Using Eq.~(2.29), we obtain
\begin{equation}
\mathcal{P}'(\rho)
=
(2s-1)\eta_s
\mathcal{A}_{s,n}^{\,2s}
\rho^{-1-\frac{2s(n-1)}{2s-1}}.
\label{2.43}
\end{equation}
 Upon integration of Eq.~(\ref{2.43}), we get
\begin{equation}
\mathcal{P}(\rho)
=
\frac{M}{(n-1)\rho^n}
-
\frac{(2s-1)^2\eta_s}
{2s(n-1)}
\mathcal{A}_{s,n}^{\,2s}
\rho^{-\frac{2s(n-1)}{2s-1}},
\label{2.45}
\end{equation}
where $M$ is an integration constant associated with the mass
parameter.

Using Eq.~(\ref{2.20}), Eq.~(\ref{2.45}) can be rearranged as
\begin{equation}
c\psi^4+\mu\psi^3+\lambda\psi^2+\psi+k(\rho)=0,
\label{2.46}
\end{equation}
where
\begin{equation}
 {
k(\rho)
=
1-\frac{M}{(n-1)\rho^n}
+ C_{s,n}
\rho^{-\frac{2s(n-1)}{2s-1}}
},
\label{2.47}
\end{equation}
where $$C_{s,n} =
\frac{(2s-1)^2\,2^{s-1}l^{\,2-2s}}
{s(n-1)^2}
\left[
\frac{s}{2q}
\left(
\frac{2q^2l^{\,n-2}}{s^2}
\right)^{\frac{s}{2s-1}}
\right]^{2s}.$$
Equation~(\ref{2.47}) is the nonlinear electromagnetic contribution to the
quartic algebraic equation. 
For $s=1$, this becomes
\begin{equation}
k(\rho)
=
1-\frac{M}{(n-1)\rho^n}
+
\frac{q^2l^{\,2n-4}}
{(n-1)^2\rho^{2n-2}},
\qquad s=1,
\label{2.49}
\end{equation}
which is the corresponding linear-Maxwell magnetic-brane expression
for the normalization adopted in the action.
It is worth noting that when $2s=n$, the exponent of the nonlinear
electromagnetic term becomes
\begin{equation}
\frac{2s(n-1)}{2s-1}=n,
\label{2.50}
\end{equation}
so that the electromagnetic contribution to $k(\rho)$ scales as
$\rho^{-n}$. The logarithmic behavior occurs in the vector potential
$A_\phi$,  rather than in $k(\rho)$.

Equation~(\ref{2.46}) is a quartic polynomial in $\psi$. To obtain its
analytic solution, we divide this equation by $c$ and introduce
\begin{equation}
\psi=y-\frac{\mu}{4c}.
\label{2.51}
\end{equation}
The quartic equation then takes the depressed form
\begin{equation}
y^4+\alpha y^2+\beta y+\gamma=0,
\label{2.52}
\end{equation}
where
\begin{eqnarray}
\alpha
&=&
-\frac{3\mu^2}{8c^2}
+\frac{\lambda}{c},
\label{2.53}\\
\beta
&=&
\frac{\mu^3}{8c^3}
-\frac{\mu\lambda}{2c^2}
+\frac{1}{c},
\label{2.54}\\
 \gamma
&=&
-\frac{3\mu^4}{256c^4}
+\frac{\lambda\mu^2}{16c^3}
-\frac{\mu}{4c^2}
+\frac{k(\rho)}{c}.
\label{2.55}
\end{eqnarray}
 The four roots of the quartic can then be represented by the standard
Ferrari expression
\begin{equation}
y
=
-\frac{5\alpha}{6}
+U-\frac{P}{3U}
+
\frac{1}{2}
\left[
\pm W
\pm
\sqrt{
-\left(3\alpha+2y_0
\pm\frac{2\beta}{W}\right)
}
\right],
\label{2.62}
\end{equation}
with
\begin{eqnarray}
P
&=&
-\frac{\alpha^2}{12}-\gamma,\ \ U
=
\left(
-\frac{H}{2}
\pm\sqrt{\Delta}
\right)^{1/3}, \ \ W
=
\sqrt{\alpha+2y_0},\label{2.56} 
\end{eqnarray}
where 
\begin{eqnarray}
H
&=&
-\frac{\alpha^3}{108}
+\frac{\alpha\gamma}{3}
-\frac{\beta^2}{8},\label{2.57}\\
 \Delta
&=&
\frac{H^2}{4}
+\frac{P^3}{27}, 
\label{2.58}\\
y_0
&=&
-\frac{5\alpha}{6}
+U-\frac{P}{3U}.
\end{eqnarray}
Here the signs are correlated according to the chosen quartic branch.
The corresponding metric function is therefore
\begin{equation}
 {
f(\rho)
=
-\frac{\rho^2}{l^2}
\left(
y-\frac{\mu}{4c}
\right).
}
\label{2.63}
\end{equation}
The physically relevant branch is selected by requiring that the metric
approach the appropriate asymptotically AdS branch and that
$f(\rho)>0$ in the physically accessible region. The lower-order
Gauss--Bonnet and cubic quasi-topological limits should be obtained
directly from the corresponding reduced polynomial rather than by
taking singular limits of the explicit quartic-root expression  \cite{29,30}.

 \section{Physical Properties of the Solutions}
\label{sec3}
In this section, we investigate the physical properties of the magnetic
brane solutions obtained in Sec.~2. In particular, we analyze the
admissible radial domain, the behavior of the curvature invariants, the
near-origin geometry after a suitable coordinate transformation, and the
associated conical singularity. We also discuss the dependence of the
solutions on the nonlinear electromagnetic parameter and the
higher-curvature coupling constants.
For the physical branch, we require
\begin{equation}
f(\rho)>0
\end{equation}
in the accessible region.

\subsection{Admissible Radial Domain}
\label{sec:admissible_domain}
A zero of the metric function is determined by
$
f(\rho_+)=0$.
Since
\begin{equation}
\psi(\rho)=-\frac{l^2 f(\rho)}{\rho^2},
\end{equation}
the condition $f(\rho_+)=0$ implies
$
\psi(\rho_+)=0.
$ which eventually gives
$
k(\rho_+)=0.
$
Consequently, the integration constant $M$ can be expressed in terms of
$\rho_+$ as
\begin{equation}
M
=
(n-1)\rho_+^n
\left[
1+
C_{s,n}\rho_+^{-\frac{2s(n-1)}{2s-1}}
\right].
\label{eq:3.7}
\end{equation}
Equation~\eqref{eq:3.7} shows explicitly that the location of the
largest zero of $f(\rho)$ does not depend explicitly on the
higher-curvature couplings $\lambda$, $\mu$, and $c$. These couplings
enter the algebraic relation determining $\psi$ and hence affect the
metric away from the lower endpoint, but they do not appear in the
condition $k(\rho_+)=0$. 
The largest real zero of $f(\rho)$ is denoted by $\rho_+$. The physical
region is then
\begin{equation}
\rho\geq \rho_+,
\label{eq:3.8}
\end{equation}
with the open region $\rho>\rho_+$ corresponding to the nonsingular
interior of the coordinate patch. The surface $\rho=\rho_+$ can not
be interpreted as an event horizon. Indeed,
\begin{equation}
g_{tt}\big|_{\rho=\rho_+}
=
-\frac{\rho_+^2}{l^2}\neq0.
\end{equation}
Thus, the zero of $f(\rho)$ does not coincide with a Killing horizon. Instead,
$\rho=\rho_+$ represents the lower endpoint of the physically
accessible radial coordinate.

For a simple zero, we require
\begin{equation}
f'(\rho_+)>0.
\label{eq:3.9}
\end{equation}
Using Eq.~\eqref{2.46}, differentiation with respect to $\rho$ gives
\begin{equation}
\left(
1+2\lambda\psi+3\mu\psi^2+4c\psi^3
\right)\psi'
+
k'(\rho)=0.
\end{equation}
At $\rho=\rho_+$, where $\psi(\rho_+)=0$, this reduces to
\begin{equation}
\psi'(\rho_+)=-k'(\rho_+).
\end{equation}
This leads to
\begin{equation}
k'(\rho)
=
\frac{nM}{(n-1)\rho^{n+1}}
-
\frac{2s(n-1)}{2s-1}
C_{s,n}
\rho^{-\frac{2s(n-1)}{2s-1}-1}. 
\end{equation}
Using Eq.~\eqref{eq:3.7}, we find
\begin{equation}
f'(\rho_+)
=
\frac{1}{l^2}
\left[
n\rho_+
+
\frac{2s-n}{2s-1}
C_{s,n}
\rho_+^{1-\frac{2s(n-1)}{2s-1}}
\right].
\label{eq:3.11}
\end{equation}
The condition $f'(\rho_+)>0$ guarantees that $f(\rho)$ becomes positive
immediately outside the lower endpoint.
It is therefore useful to introduce the coordinate transformation
\begin{equation}
r=\sqrt{\rho^2-\rho_+^2},
\qquad
\rho=\sqrt{r^2+\rho_+^2}.
\label{eq:3.12}
\end{equation}
The physical region $\rho\geq\rho_+$ is mapped onto
$
0\leq r<\infty.$
The endpoint $\rho=\rho_+$ is now represented by $r=0$.
Under the transformation \eqref{eq:3.12}, the metric becomes
\begin{equation}
ds^2
=
-\frac{r^2+\rho_+^2}{l^2}dt^2
+
\frac{r^2}
{(r^2+\rho_+^2)f(r)}\,dr^2
+
l^2 f(r)d\phi^2
+
\frac{r^2+\rho_+^2}{l^2}dX^2,
\label{eq:3.14}
\end{equation}
where
\begin{equation}
f(r)
\equiv
f\left(\sqrt{r^2+\rho_+^2}\right).
\end{equation}
Because $f(\rho_+)=0$, one has
\begin{equation}
f(0)=0.
\label{eq:3.15}
\end{equation}
Furthermore,
\begin{equation}
\left.\frac{df(r)}{dr}\right|_{r=0}=0,
\label{eq:3.16}
\end{equation}
and the first nonvanishing term in the expansion of $f(r)$ is quadratic
in $r$.
 
Now, using
\begin{equation}
\frac{d\rho}{dr}
=
\frac{r}{\sqrt{r^2+\rho_+^2}},
\end{equation}
the nonvanishing component in the new radial coordinate is
\begin{equation}
F_{\phi r}
=
F_{\phi\rho}\frac{d\rho}{dr}.
\end{equation}
Hence,
\begin{equation}
F_{\phi r}
=
\mathcal{A}_{s,n}\,
r\,
(r^2+\rho_+^2)^{-\frac{1}{2}
\left(
1+\frac{n-1}{2s-1}
\right)}.
\label{eq:3.19}
\end{equation}
It follows immediately that
\begin{equation}
F_{\phi r}\big|_{r=0}=0,
\end{equation}
and the electromagnetic field is finite at the endpoint $r=0$.
The corresponding function $k(r)$ is obtained becomes
\begin{equation}
k(r)
=
1
-
\frac{M}
{(n-1)(r^2+\rho_+^2)^{n/2}}
+
C_{s,n}
(r^2+\rho_+^2)^{-\frac{s(n-1)}{2s-1}},
\label{eq:3.21}
\end{equation}
which vanishes at $r=0$.

\subsection{Curvature Singularity and Kretschmann Scalar}

To examine the curvature properties of the spacetime, we consider the
Kretschmann scalar,
\begin{equation}
\mathcal{K}
=
R_{\mu\nu\alpha\beta}
R^{\mu\nu\alpha\beta}.
\end{equation}
For the metric considered here, it can be written as
\begin{equation}
\mathcal{K}
=
f''(\rho)^2
+
\frac{2(n-1)}{\rho^2}f'(\rho)^2
+
\frac{2(n-1)(n-2)}{\rho^4}f(\rho)^2,
\label{eq:3.22}
\end{equation}
where primes denote differentiation with respect to $\rho$.

The expression \eqref{eq:3.22} can become singular as
$\rho\rightarrow0$. However, the coordinate value $\rho=0$ is not
contained in the physical domain of the magnetic brane. The physical
region satisfies
\begin{equation}
\rho\geq\rho_+>0.
\end{equation}
Therefore, all curvature invariants remain finite throughout the
physically accessible region, provided that the selected branch of the
quartic solution is regular there.
Consequently, the magnetic brane does not possess a curvature
singularity in its physical domain. Nevertheless, as we show below,
the endpoint $r=0$ has a nontrivial conical structure.

\subsection{Near-Endpoint Geometry and Conical Singularity}

To determine the nature of the endpoint $r=0$, we expand the metric
function around this point. From
\begin{equation}
\rho-\rho_+
=
\sqrt{r^2+\rho_+^2}-\rho_+
=
\frac{r^2}{2\rho_+}
+\mathcal{O}(r^4),
\end{equation}
and
\begin{equation}
f(\rho)
=
f'(\rho_+)(\rho-\rho_+)
+\mathcal{O}\left((\rho-\rho_+)^2\right),
\end{equation}
we obtain
\begin{equation}
f(r)
=
\frac{f'(\rho_+)}{2\rho_+}r^2
+\mathcal{O}(r^4).
\label{eq:3.23}
\end{equation}
Therefore, the $(r,\phi)$ sector of the metric near $r=0$ becomes
\begin{equation}
ds^2_{(r,\phi)}
\simeq
\frac{2}{\rho_+f'(\rho_+)}dr^2
+
\frac{l^2f'(\rho_+)}{2\rho_+}r^2d\phi^2.
\label{eq:3.24}
\end{equation}
Introducing a proper radial coordinate
$
R
=
\sqrt{\frac{2}{\rho_+f'(\rho_+)}}\,r,
$
the two-dimensional transverse metric takes the form
\begin{equation}
ds^2_{(r,\phi)}
\simeq
dR^2
+
\frac{l^2f'(\rho_+)^2}{4}R^2d\phi^2.
\label{eq:3.25}
\end{equation}
Since $\phi$ has period $2\pi$, the ratio of the circumference to the
proper radius is
\begin{equation}
\lim_{R\rightarrow0}
\frac{\mathcal{C}}{R}
=
\pi l f'(\rho_+).
\label{eq:3.26}
\end{equation}
Thus, the geometry is regular at the endpoint only if
$
{l f'(\rho_+)}/2=1.
$
In general this condition is not satisfied, and the endpoint possesses
a conical singularity.
The angular deficit is therefore
\begin{equation}
\delta_\phi
=
2\pi
-
\lim_{R\rightarrow0}\frac{\mathcal{C}}{R},
\end{equation}
or explicitly,
\begin{equation}
 {
\delta_\phi
=
2\pi
\left[
1-\frac{l f'(\rho_+)}{2}
\right].
}
\label{eq:3.27}
\end{equation}
It is convenient to introduce the deficit parameter $\tau$ through
\begin{equation}
\delta_\phi=8\pi\tau.
\label{eq:3.28}
\end{equation}
Consequently,
\begin{equation}
 {
\tau
=
\frac14
\left[
1-\frac{l f'(\rho_+)}{2}
\right].
}
\label{eq:3.29}
\end{equation}
Using Eq.~\eqref{eq:3.11}, the deficit angle can be expressed as
\begin{equation}
\delta_\phi
=
2\pi
\left\{
1
-
\frac{1}{2l}
\left[
n\rho_+
+
\frac{2s-n}{2s-1}
C_{s,n}
\rho_+^{1-\frac{2s(n-1)}{2s-1}}
\right]
\right\}.
\label{eq:3.30}
\end{equation}
This result demonstrates an important property of the solution:
$\delta_\phi$ does not contain the higher-curvature couplings
$\lambda$, $\mu$, and $c$ explicitly. Its dependence on the
electromagnetic sector enters through $C_{s,n}$ and the endpoint
$\rho_+$.

For a genuine positive angular deficit, one requires $
0<\delta_\phi<2\pi,
$
which corresponds to
\begin{equation}
0<
\frac{l f'(\rho_+)}{2}
<1.
\label{eq:3.31}
\end{equation}
If $l f'(\rho_+)/2=1$, the conical defect disappears and values outside
this range correspond to an excess angle rather than an ordinary
deficit.

\subsection{Numerical Behavior of the Metric Function}

We now discuss the qualitative behavior of the metric function using
the numerical solutions. 
The numerical
analysis is performed using the same branch-selection
conditions specified after Eq.~(2.63), namely the asymptotically AdS
branch satisfying
\begin{equation}
f(\rho)>0
\end{equation}
throughout the physical domain.

\begin{figure}[!htb]
\centering
 {
\includegraphics[width=0.70\textwidth]{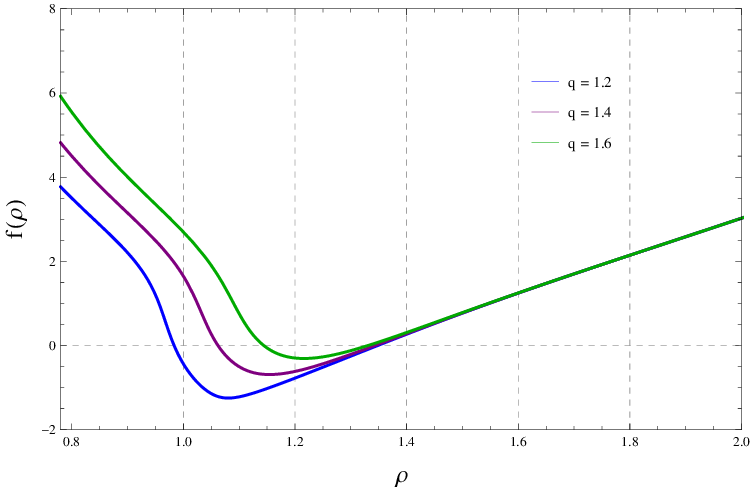}
}
\caption{{Metric function $f(\rho)$ for the Einstein/GR-connected branch of the quartic quasi-topological magnetic brane coupled to power-law Maxwell nonlinear electrodynamics, for $M=10$, $s=0.6$, $n=4$, $\lambda=0.01$, $\mu=0.02$, $c=-0.1$, and $l=1$. The curves correspond to the magnetic charge parameters $q=1.2$, $1.4$, and $1.6$. }}
\label{Fig1}
\end{figure}

Fig.~ \ref{Fig1} illustrates the numerical behavior of the metric function for
different values of the magnetic parameter $q$. For the parameter choices
shown, $f(\rho)$ may possess more than one real zero; however, the physical
solution is defined on the branch satisfying $f(\rho)>0$ for
$\rho\geq\rho_{+}$, where $\rho_{+}$ denotes the largest real zero.
The surface $\rho=\rho_{+}$ is not a Killing horizon because
$g_{tt}(\rho_{+})\neq0$; instead, it represents the lower endpoint of the
physically accessible radial coordinate. Increasing $q$ changes the
electromagnetic contribution to $k(\rho)$ and consequently shifts
$\rho_{+}$.

\begin{figure}[!htb]
\centering{
    \includegraphics[width=0.70\textwidth]{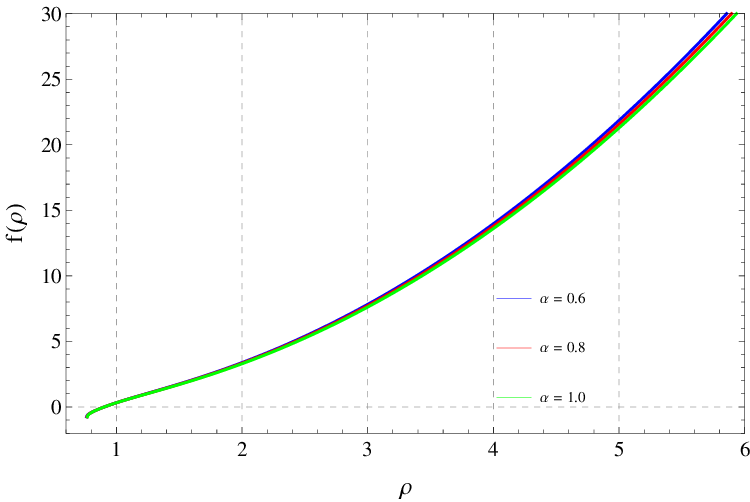}\label{Fig2} }
   \caption{Metric function $f(\rho)$ for the Einstein/GR-connected branch of the quartic quasi-topological magnetic brane coupled to power-law Maxwell nonlinear electrodynamics for $M=2$, $s=0.9$, $q=0.001$, $\lambda=-0.001$, $n=4$, $c=-0.1$, and $l=1$. The dependence of the metric function on the parameter $\alpha$ is shown for $\alpha=0.6$, $0.8$, and $1.0$. }
    
    \end{figure}

   In Fig. \ref{Fig2}, the three curves remain close at small $\rho$ and become increasingly distinguishable at larger radial distances, illustrating the effect of the quartic-gravity parameter $\alpha$ on the radial structure of the magnetic-brane geometry.
    
\begin{figure}[!htb]
   \label{fig3} 
   \centering{ \includegraphics[width=0.7\textwidth]{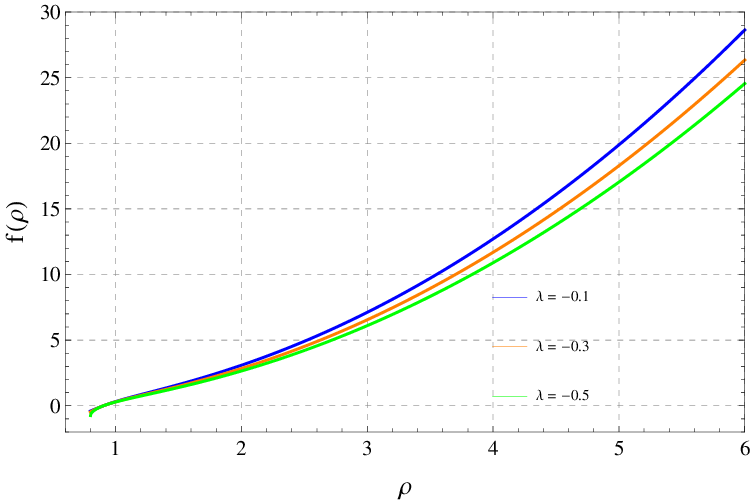}}  %
\caption{Metric function $f(\rho)$ for the Einstein/GR-connected branch of the quartic quasi-topological magnetic brane coupled to power-law Maxwell nonlinear electrodynamics. The parameters are fixed at $M=2$, $s=0.9$, $q=0.001$, $\mu=0.2$, $n=4$, $c=-0.1$, and $l=1$, while the quartic quasi-topological coupling is varied as $\lambda=-0.1,-0.3,-0.5$. }  
    \end{figure}

  The Fig. \ref{fig3} illustrates the dependence of the metric function on the coupling parameter $\lambda$, with an increasingly pronounced separation among the curves at larger radial distances $\rho$.
    
\begin{figure}[htb] 
 \centering{   \includegraphics[width=0.7\textwidth]{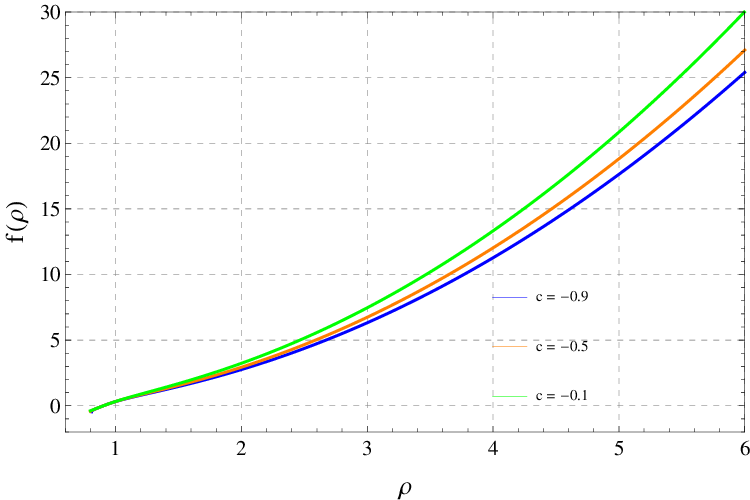}%
\caption{Metric function $f(\rho)$ for the Einstein/GR-connected branch of the quartic quasi-topological magnetic brane coupled to power-law Maxwell nonlinear electrodynamics. The parameters are fixed at $M=2$, $s=0.9$, $q=0.001$, $\mu=0.2$, $n=4$, $\lambda=-0.001$, and $l=1$, while the quartic quasi-topological coupling is varied as $c=-0.9$, $-0.5$, and $-0.1$. } \label{fig4}%
}
\end{figure}

     The Fig.  \ref{fig4} demonstrates the dependence of the metric function on the quartic coupling $c$, with the separation between the curves becoming more pronounced at larger radial distances $\rho$.  
For fixed values of $M$, $s$, $n$, $\lambda$, $\mu$, and $c$, variation
of the magnetic parameter $q$ changes the electromagnetic contribution
to $k(\rho)$ and consequently changes the location of the largest root
$\rho_+$. Thus, the magnetic parameter controls the size of the
excluded region
\begin{equation}
0\leq\rho<\rho_+.
\end{equation}

On the other hand, the higher-curvature parameters $\lambda$, $\mu$,
and $c$ do not enter the root condition
$k(\rho_+)=0$. Therefore, when the mass parameter is re-expressed in
terms of $\rho_+$ using Eq.~\eqref{eq:3.7}, the endpoint $\rho_+$ is
independent of these couplings. Their influence becomes apparent in
the solution for $\psi(\rho)$ and hence in the detailed behavior of
$f(\rho)$ away from $\rho=\rho_+$.

Near the lower endpoint, the expansion
\begin{equation}
f(\rho)
=
f'(\rho_+)(\rho-\rho_+)
+\mathcal{O}\left((\rho-\rho_+)^2\right)
\end{equation}
shows that the leading behavior is controlled by
$f'(\rho_+)$, which depends on the electromagnetic parameters through
$C_{s,n}$ and $\rho_+$. The higher-curvature coefficients enter only at
higher orders through the nonlinear algebraic relation between
$f(\rho)$ and $\psi(\rho)$.

At large $\rho$, the electromagnetic and mass contributions in
$k(\rho)$ decay as $
\rho^{-n}
$
and
$
\rho^{-\frac{2s(n-1)}{2s-1}},
$ 
respectively. For
$
s>\frac12,
$
the nonlinear electromagnetic contribution decreases at large
$\rho$. Consequently, the asymptotic behavior is governed by the
vacuum quartic equation
\begin{equation}
c\psi_\infty^4
+
\mu\psi_\infty^3
+
\lambda\psi_\infty^2
+
\psi_\infty
+
1
=0.
\label{eq:3.32}
\end{equation}
The physically relevant root of Eq.~\eqref{eq:3.32} determines the
effective asymptotic AdS scale and must be chosen consistently with the
branch condition discussed in Sec.~\ref{sec2}.

Thus, the numerical behavior of $f(\rho)$ can be summarized as
follows. The magnetic parameters predominantly control the position of
the lower endpoint and the near-endpoint geometry, whereas the
higher-curvature parameters determine the asymptotic branch and the
large-$\rho$ profile of the metric function. The conical structure at
$r=0$ is encoded in Eq.~\eqref{eq:3.27}.
 \subsection{Conical Singularity and Deficit Angle}
\label{sec:conical_singularity}

As discussed in subsection \ref{sec:admissible_domain}, the physical region of the
magnetic brane is restricted to $\rho\geq \rho_{+}$, where $\rho_{+}$ denotes
the largest real zero of the metric function,
\begin{equation}
f(\rho_{+})=0.
\end{equation}
This surface does not represent an event horizon, since
$g_{tt}|_{\rho=\rho_{+}}\neq 0$. Instead, it corresponds to the lower endpoint
of the physically accessible radial coordinate. The coordinate transformation
\begin{equation}
r=\sqrt{\rho^{2}-\rho_{+}^{2}},
\qquad
\rho=\sqrt{r^{2}+\rho_{+}^{2}},
\label{eq:rho_to_r}
\end{equation}
maps the physical region $\rho\geq\rho_{+}$ onto $0\leq r<\infty$, with
$r=0$ corresponding to $\rho=\rho_{+}$.

The curvature analysis presented in the previous subsection shows that the
Kretschmann scalar remains finite throughout the physical region, provided
that the physically relevant branch of the quartic equation is selected.
Thus, the endpoint $r=0$ does not correspond to a curvature singularity.
Nevertheless, the geometry at this endpoint may possess a conical
singularity. To examine this possibility, we expand the metric function
near $r=0$. From Eq.~\eqref{eq:rho_to_r}, one obtains
\begin{equation}
\rho-\rho_{+}
=
\sqrt{r^{2}+\rho_{+}^{2}}-\rho_{+}
=
\frac{r^{2}}{2\rho_{+}}+\mathcal{O}(r^{4}).
\label{eq:rho_expansion}
\end{equation}
Since $f(\rho_{+})=0$, the Taylor expansion of $f(\rho)$ around
$\rho=\rho_{+}$ gives
\begin{equation}
f(\rho)
=
f'(\rho_{+})(\rho-\rho_{+})
+\mathcal{O}\left((\rho-\rho_{+})^{2}\right).
\label{eq:f_rho_expansion}
\end{equation}
Consequently, near $r=0$,
\begin{equation}
f(r)
=
\frac{f'(\rho_{+})}{2\rho_{+}}r^{2}
+\mathcal{O}(r^{4}).
\label{eq:f_r_expansion}
\end{equation}
Using this result in the $(r,\phi)$ sector of the metric, we find that
\begin{equation}
ds^{2}_{(r,\phi)}
\simeq
\frac{2}{\rho_{+}f'(\rho_{+})}\,dr^{2}
+
\frac{l^{2}f'(\rho_{+})}{2\rho_{+}}r^{2}d\phi^{2}.
\label{eq:near_origin_metric}
\end{equation}
Introducing the proper radial coordinate
$
R=
\sqrt{\frac{2}{\rho_{+}f'(\rho_{+})}}\,r,
$
the transverse two-dimensional metric takes the form
\begin{equation}
ds^{2}_{(R,\phi)}
\simeq
dR^{2}
+
\frac{l^{2}f'(\rho_{+})^{2}}{4}
R^{2}d\phi^{2}.
\label{eq:transverse_metric}
\end{equation}
Since the angular coordinate has the standard periodicity
$0\leq\phi\leq2\pi$, the ratio of the circumference to the proper radius
in the limit $R\rightarrow0$ is
\begin{equation}
\lim_{R\rightarrow0}\frac{C}{R}
=
\pi l f'(\rho_{+}).
\label{eq:circumference_radius}
\end{equation}
Therefore, the endpoint $r=0$ is regular only if
\begin{equation}
\frac{l f'(\rho_{+})}{2}=1.
\label{eq:regularity_condition}
\end{equation}
In general, this condition is not satisfied, indicating that the spacetime
possesses a conical singularity at $r=0$ despite the absence of a curvature
singularity.
The corresponding angular deficit is defined by
\begin{equation}
\delta\phi
=
2\pi-
\lim_{R\rightarrow0}\frac{C}{R},
\end{equation}
which yields
\begin{equation}
 {
\delta_\phi
=
2\pi
\left[
1-\frac{l f'(\rho_{+})}{2}
\right].
}
\label{eq:deficit_angle}
\end{equation}
It is convenient to introduce the dimensionless deficit parameter $\tau$ by
\begin{equation}
\delta_\phi=8\pi\tau,
\label{eq:tau_definition}
\end{equation}
so that
\begin{equation}
 {
\tau
=
\frac{1}{4}
\left[
1-\frac{l f'(\rho_{+})}{2}
\right].
}
\label{eq:tau}
\end{equation}
Using Eq.~\eqref{eq:3.11} for the derivative of the metric function at
the lower endpoint, the deficit angle can alternatively be written as
\begin{equation}
\delta_\phi
=
2\pi
\left\{
1-
\frac{1}{2l}
\left[
n\rho_{+}
+
\frac{2s-n}{2s-1}
C_{s,n}
\rho_{+}^{\,1-\frac{2s(n-1)}{2s-1}}
\right]
\right\},
\label{eq:deficit_explicit}
\end{equation}
where
\begin{equation}
C_{s,n}
=
\frac{(2s-1)^{2}2^{s-1}l^{2-2s}}
{s(n-1)^{2}}
\left[
\frac{s}{2q}
\left(
\frac{2q^{2}l^{n-2}}{s^{2}}
\right)^{\frac{s}{2s-1}}
\right]^{2s}.
\label{eq:Csn_deficit}
\end{equation}
This expression shows explicitly that the higher-curvature coefficients
$\lambda$, $\mu$, and $c$ do not appear explicitly in the deficit-angle
formula. Their effects enter the metric function through the quartic
algebraic equation, whereas the derivative $f'(\rho_{+})$ at the endpoint is
determined by the electromagnetic contribution and $\rho_{+}$.

\begin{figure}[ht]
\centering
{
\includegraphics[width=0.7\textwidth]{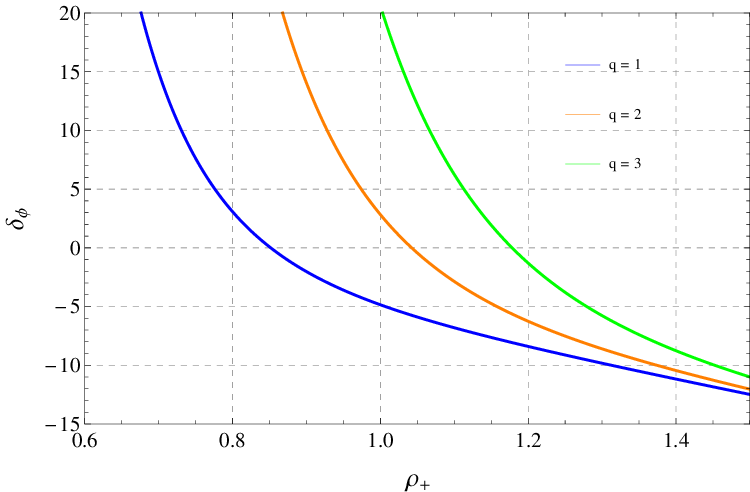}
}
\caption{Angular deficit $\delta_{\phi}$ as a function of the endpoint parameter $\rho_{+}$ for the power-law Maxwell nonlinear electrodynamics with $s=0.8$ and $n=4$. The curves correspond to different values of the magnetic charge parameter, $q=1$, $2$, and $3$, with $l=1$. }
\label{Fig5}
\end{figure}

\begin{figure}[!htb]
\centering
{
\includegraphics[width=0.7\textwidth]{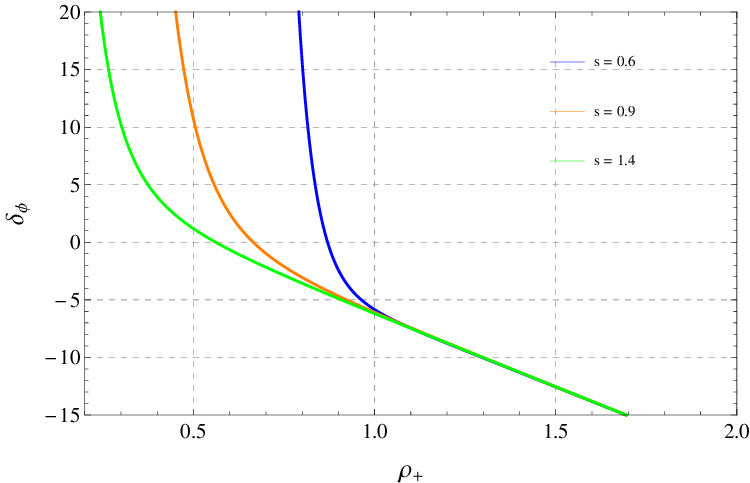}
}
\caption{Angular deficit $\delta_{\phi}$ as a function of the endpoint parameter $\rho_{+}$ for fixed magnetic charge $q=0.5$ and spacetime dimension parameter $n=4$, with $l=1$. The curves correspond to different values of the power-law Maxwell nonlinear parameter, $s=0.6$, $0.9$, and $1.4$.}
\label{Fig6}
\end{figure}

In Fig. \ref{Fig5}, we see that the  deficit angle decreases with increasing $\rho_{+}$, while increasing $q$ shifts the zero-deficit point toward larger values of $\rho_{+}$. The physical conical-deficit region is characterized by $0<\delta_{\phi}<2\pi$.
In Fig. \ref{Fig6}, we see that the angular deficit decreases with increasing $\rho_{+}$, while increasing $s$ shifts the zero-deficit point toward smaller values of $\rho_{+}$. The physical conical-deficit region is characterized by $0<\delta_{\phi}<2\pi$.

For a genuine conical deficit, the angular deficit must satisfy
$
0<\delta_\phi<2\pi,
\label{eq:positive_deficit}
$
or equivalently,
\begin{equation}
0<
\frac{l f'(\rho_{+})}{2}
<1.
\label{eq:deficit_condition}
\end{equation}
For  $
\frac{l f'(\rho_{+})}{2}=1,
$
the conical defect disappears and the geometry becomes locally regular at
the endpoint. Values outside the above interval correspond to an angular
excess rather than an ordinary deficit.

The behavior of the deficit angle is therefore controlled primarily by the
magnetic parameter $q$, the nonlinear parameter $s$, and the endpoint
$\rho_{+}$. In particular, the dependence on the higher-curvature parameters
$\lambda$, $\mu$, and $c$ is not explicit in Eq.~\eqref{eq:deficit_explicit}.
This result provides a useful distinction between the near-endpoint geometry
and the asymptotic behavior of the magnetic brane: the former is governed
predominantly by the electromagnetic sector, whereas the latter is sensitive
to the higher-curvature couplings through the asymptotic branch of the
quartic equation.
\section{Conserved Quantities of a Magnetic Static Brane}
\label{sec4}

As discussed in Sec.~\ref{sec3}, the static magnetic-brane solution is
horizonless. The surface $\rho=\rho_+$, or equivalently $r=0$ after the
coordinate transformation introduced in Eq.~\eqref{eq:3.12}, is the
lower endpoint of the physically accessible radial region rather than
an event horizon. Consequently, the usual black-hole thermodynamic
quantities associated with a horizon are not introduced here. Instead,
we determine the conserved quantities of the spacetime, in particular
the mass and electric charge densities.

For asymptotically AdS solutions, the gravitational action must be
supplemented by appropriate boundary terms in order to obtain a
well-defined variational principle. We therefore consider the action
\begin{equation}
I_1=I_G+I_b,
\qquad
I_b=
I_b^{(1)}+I_b^{(2)}+I_b^{(3)}+I_b^{(4)},
\label{eq:4.1}
\end{equation}
where $I_b$ denotes the total boundary contribution. The terms
$I_b^{(1)}$, $I_b^{(2)}$, $I_b^{(3)}$, and $I_b^{(4)}$ correspond,
respectively, to the boundary terms associated with the
Einstein--Hilbert, Gauss--Bonnet, third-order quasi-topological, and
fourth-order quasi-topological sectors
\cite{32,33,34,35,36,37,38}.

Although the action \eqref{eq:4.1} provides the appropriate boundary
terms for the variational problem, the corresponding on-shell action
and boundary stress tensor contain divergences when evaluated at the
AdS boundary. Following the standard counterterm prescription
\cite{31,39,40}, we therefore supplement the action with a local
counterterm. For the present flat boundary geometry,
\begin{equation}
\widehat{R}_{abcd}(\gamma)=0,
\end{equation}
the required counterterm takes the simple form
\begin{equation}
I_{\rm ct}
=
\frac{1}{8\pi}
\int_{\partial\mathcal{M}}
d^n x\,\sqrt{-\gamma}\,
\frac{n-1}{l_{\rm eff}},
\label{eq:4.2}
\end{equation}
where $l_{\rm eff}$ denotes the effective AdS curvature radius of the
asymptotic vacuum branch. It is determined by the asymptotic value
$\psi_\infty$ of the metric function through
\begin{equation}
\psi_\infty
=
-\frac{l^2}{l_{\rm eff}^{\,2}},
\label{eq:4.3}
\end{equation}
and therefore satisfies the vacuum quartic equation
\begin{equation}
c\psi_\infty^4
+\mu\psi_\infty^3
+\lambda\psi_\infty^2
+\psi_\infty
+1=0.
\label{eq:4.4}
\end{equation}
The physically relevant root is the one corresponding to the
asymptotically AdS branch selected in Sec.~2. In the Einstein limit
$\lambda=\mu=c=0$, Eq.~\eqref{eq:4.4} gives
$\psi_\infty=-1$, and consequently
\begin{equation}
l_{\rm eff}=l.
\end{equation}

The finite action used in the calculation of the conserved quantities
is therefore
\begin{equation}
I_{\rm ren}
=
I_G+I_b+I_{\rm ct}.
\label{eq:4.5}
\end{equation}

\subsection{Boundary Stress Tensor and Conserved Mass}

To construct the quasi-local conserved quantities, we consider a
timelike boundary $\partial\mathcal{M}$ at large radial coordinate.
Let $\mathcal{B}$ be a spacelike $(n-1)$-dimensional section of this
boundary, with induced metric $\sigma_{ij}$. The boundary metric can
be written in the ADM form
\begin{equation}
\gamma_{ab}dx^a dx^b
=
-\mathcal{N}^2dt^2
+
\sigma_{ij}
\left(d\phi^i+V^i dt\right)
\left(d\phi^j+V^jdt\right),
\label{eq:4.6}
\end{equation}
where $\mathcal{N}$ and $V^i$ are, respectively, the boundary lapse
and shift functions. The notation $\mathcal{N}$ is used here to
distinguish the boundary lapse from the bulk function $N(\rho)$
introduced in Sec.~\ref{sec2}, which has already been fixed to unity on the
nondegenerate branch.

The renormalized boundary stress tensor is obtained by variation of
the finite action \eqref{eq:4.5} with respect to the boundary metric:
\begin{equation}
T^{ab}
=
\frac{2}{\sqrt{-\gamma}}
\frac{\delta I_{\rm ren}}
{\delta\gamma_{ab}}.
\label{eq:4.7}
\end{equation}
For any Killing vector $\xi^a$ of the boundary geometry, the associated
quasi-local conserved quantity is
\begin{equation}
\mathcal{Q}(\xi)
=
\int_{\mathcal{B}}
d^{\,n-1}\phi\,
\sqrt{\sigma}\,
T_{ab}u^a\xi^b,
\label{eq:4.8}
\end{equation}
where $u^a$ is the timelike unit normal to $\mathcal{B}$ within
$\partial\mathcal{M}$.

For the static magnetic brane, the relevant Killing vector associated
with time translations is
\begin{equation}
\xi^a=\left(\frac{\partial}{\partial t}\right)^a.
\label{eq:4.9}
\end{equation}
Evaluating Eq.~\eqref{eq:4.8} at the asymptotic boundary and using the
normalization of the integration constant $M$ adopted in Sec.~2 gives
the mass density
\begin{equation}
 {
M_{\rm total}
=
\frac{M}{4(n-1)}.
}
\label{eq:4.10}
\end{equation}
Here, $M_{\rm total}$ denotes the conserved mass per unit transverse
volume, with the overall normalization corresponding to the
gravitational conventions adopted in the action. Since the
counterterm \eqref{eq:4.2} cancels the divergent contributions at
large radius, the resulting conserved mass is finite.

It is important to emphasize that $M$ in Eq.~\eqref{eq:4.10} is the
same integration constant appearing in the quartic equation
\eqref{2.46}. In particular, when the lower endpoint is parametrized
by $\rho_+$, it is given by
\begin{equation}
M
=
(n-1)\rho_+^n
\left[
1+
C_{s,n}
\rho_+^{-\frac{2s(n-1)}{2s-1}}
\right].
\label{eq:4.11}
\end{equation}
 Thus, the conserved
mass can equivalently be written as
\begin{equation}
M_{\rm total}
=
\frac{\rho_+^n}{4}
\left[
1+
C_{s,n}
\rho_+^{-\frac{2s(n-1)}{2s-1}}
\right].
\label{eq:4.12}
\end{equation}

\subsection{Electric Field and Electric Charge}

Although the solution possesses a nonvanishing magnetic field, it does
not carry a net electric charge. This can be shown directly from the
form of the electromagnetic field obtained in Sec.~3.

After the transformation
\begin{equation}
\rho=\sqrt{r^2+\rho_+^2},
\end{equation}
the only nonvanishing component of the field-strength tensor is
\begin{equation}
F_{\phi r}
=
\mathcal{A}_{s,n}\,
r\,
(r^2+\rho_+^2)^{-\frac{2s+n-2}{2(2s-1)}}.
\label{eq:4.13}
\end{equation}
Thus, $
F_{tr}=
F_{t\phi}=0,$ 
and there is no electric component of the field.

To make this explicit, we consider the timelike unit vector normal to a
constant-time hypersurface
\begin{equation}
u^a
=
\frac{1}{\mathcal{N}}
\left(
\frac{\partial}{\partial t}
\right)^a
\label{eq:4.15}
\end{equation}
for the static configuration. The electric field measured by these
observers is 
\begin{equation}
E_a
=
F_{ab}u^b,
\label{eq:4.16}
\end{equation}
Since the only nonzero component of $F_{ab}$ is $F_{\phi r}$ and the
observer has only a time component, it follows immediately that
\begin{equation}
E_a=0.
\label{eq:4.18}
\end{equation}
Therefore, the electric flux through an asymptotic hypersurface
vanishes:
\begin{equation}
\mathcal{Q}_{e}
=
\frac{1}{4\pi}
\int_{\partial\mathcal{B}}
{}^{\star}F
=
0.
\label{eq:4.19}
\end{equation}
Hence, the electric charge density of the static magnetic brane is
\begin{equation}
 {
Q_e=0.
}
\label{eq:4.20}
\end{equation}
This result is a direct consequence of the static magnetic ansatz. The
solution contains a longitudinal magnetic field but no electric field
component. Therefore, the magnetic parameter $q$  should not be interpreted as an
electric charge. A nonzero electric charge would require an electric
component of the gauge field, which is absent for the static ansatz
considered here.

In summary, the conserved quantities of the static magnetic brane are
characterized by a finite mass density,
\begin{equation}
M_{\rm total}=\frac{M}{4(n-1)},
\end{equation}
while its net electric charge, $
Q_e=0$, vanishes. 
The magnetic sector is instead encoded in the parameter $q$ through the
nonlinear magnetic field.
\section{Concluding Results}\label{sec5}

In this work, we have investigated static magnetic brane solutions in
$(n+1)$-dimensional quartic quasi-topological gravity coupled to power-law
Maxwell nonlinear electrodynamics. We have extended previous studies of
magnetic branes in higher-curvature gravity coupled to Born--Infeld,
exponential, and logarithmic nonlinear electrodynamics. By employing a
static magnetic ansatz, we have derived an algebraic equation for the metric
function and have analyzed the resulting geometry, curvature properties,
conical structure, and conserved quantities.

We have obtained horizonless magnetic brane solutions with finite curvature
invariants throughout the physically accessible region. We have shown that
the largest zero of the metric function, $\rho_{+}$, does not represent an
event horizon but instead defines the lower endpoint of the physical radial
coordinate. We have also demonstrated that the position of this endpoint
does not depend explicitly on the higher-curvature couplings
$\lambda$, $\mu$, and $c$, while it depends on the magnetic parameter $q$.
Furthermore, we have found that the higher-curvature couplings modify the
radial behavior of the metric function, particularly away from the lower
endpoint and in the asymptotic region. 

We have found that, despite the finiteness of the curvature invariants in
the physical region, the geometry generally develops a conical singularity
at the endpoint. We have shown that the associated angular deficit is
primarily controlled by the magnetic parameter $q$, the nonlinear parameter
$s$, and the endpoint $\rho_{+}$. Our analysis has demonstrated that the
deficit angle depends on $q$ and $s$ over an intermediate range of
$\rho_{+}$, whereas this dependence becomes suppressed for sufficiently
large $\rho_{+}$. We have further established that the higher-curvature
couplings do not appear explicitly in the deficit-angle expression; their
effects enter indirectly through the metric function and its algebraic
branch structure.  

We have also analyzed the asymptotic behavior of the solutions and have
shown that the nonlinear electromagnetic contribution decreases at large
radial distances for $s>1/2$. Consequently, we have found that the
asymptotic geometry is governed by the vacuum quartic equation associated
with the higher-curvature gravitational sector. Our results have therefore
shown that the magnetic and nonlinear-electromagnetic parameters primarily
control the near-endpoint geometry, whereas the quasi-topological couplings
determine the asymptotic branch and the large-distance behavior of the
spacetime.

Using the counterterm method, we have calculated the conserved mass density
and have found it to be finite. We have also demonstrated that the electric
field vanishes identically for the static magnetic configuration and,
consequently, that the net electric charge vanishes. Thus, we have
identified $q$ as the parameter characterizing the magnetic sector rather
than an electric charge for the static solution considered here.  

We have further shown that the power-law Maxwell model provides a controlled
deformation of Maxwell electrodynamics. In particular, we have recovered
linear Maxwell theory exactly in the limit $s=1$, whereas the
Born--Infeld, logarithmic, and exponential models generally recover
Maxwell electrodynamics only in the appropriate weak-field limit. We have
also recovered Einstein gravity by setting the higher-curvature couplings
to zero, $\lambda=\mu=c=0$. These limits have demonstrated that the present
model provides a useful framework for studying the interplay between
nonlinear electrodynamics and higher-curvature gravitational corrections.

Finally, we have identified several directions for future investigation.
We have planned to extend the present analysis to rotating magnetic brane
solutions, where a nonvanishing electric field and electric charge may
emerge. We have also planned to investigate the thermodynamic properties,
stability, critical behavior, and possible phase transitions of these
solutions. In addition, we intend to extend the analysis to black-hole
configurations with roots of the metric function. Such extensions will allow us
to examine more systematically the interplay between horizon structure,
quartic quasi-topological corrections, and power-law nonlinear
electrodynamics.

\section*{Acknowledgements} 
This research was funded by the Science Committee of the Ministry of Science and Higher Education of the Republic of Kazakhstan (Grant No. AP23487178).  

\end{document}